\documentclass[pra,a4paper,showpacs,superscriptaddress]{revtex4}
\usepackage{amsmath}
\usepackage{amsfonts}
\usepackage{graphicx}
\usepackage{longtable}
\newcommand{\be}{\begin{equation}}
\newcommand{\ee}{\end{equation}}
\newcommand{\bea}{\begin{eqnarray}}
\newcommand{\eea}{\end{eqnarray}}

\newcommand{\ba}[1]{\left(\begin{array}{#1}}
\newcommand{\ea}{\end{array}\right)}
\begin{document}

\title{Monogamous nature of symmetric N-qubit states of the W-class: Concurrence and Negativity Tangle} 
\author{P. J. Geetha}
\affiliation{Department of Physics, Kuvempu University, 
Shankaraghatta, Shimoga-577 451, India}
\author{K.O. Yashodamma}
\affiliation{Department of Physics, Kuvempu University, 
Shankaraghatta, Shimoga-577 451, India}
\author{Sudha}
\email{arss@rediffmail.com}
\affiliation{Department of Physics, Kuvempu University, 
Shankaraghatta, Shimoga-577 451, India}
\affiliation{Inspire Institute Inc., Alexandria, Virginia, 22303, USA.}
\date{\today}
\begin{abstract} 
Using Majorana representation of symmetric $N$-qubit pure states, we have examined the monogamous nature of the family of states with two-distinct spinors, the W-class of states. We have evaluated the $N$-concurrence tangle and showed that all the states in this family have vanishing concurrence tangle. The negativity tangle for the W-class of states is shown to be non-zero illustrating the fact that concurrence tangle underestimates the residual entanglement in a pure $N$-qubit state.
\end{abstract}

\maketitle

\section{Introduction}
Monogamy of quantum correlations/entanglement, the quantum mechanical feature indicating the restricted shareability of quantum correlations/entanglement among several parties of a composite system, has evoked a lot of interest in the recent years${[1-28]}$.  The pioneering work of Coffman, Kundu and Wootters~\cite{ckw} has led to a plethora of activity including issues such as monogamy of quantum versus classical correlations~\cite{kw2,gl}, 
monogamy  using generalized entropies\cite{bs2,renyi,bs3} and monogamy of quantum correlations other than entanglement\cite{seevinck,prabhu,sudha,xi,hf2,bruss,sqd,salini,pjg1}.   
Quantifying multi-party entanglement is another important issue and measures such as three-tangle (or concurrence-tangle)\cite{ckw} and negativity tangle\cite{neg}, based on monogamy inequality, have been proposed for quantifying {\emph {residual}} or {\emph {three-party}} entanglement in $3$-qubit pure states. In fact, {\emph {residual entanglement in a $3$-qubit state is defined as the entanglement between the qubits that is not accounted for by the two-qubit entanglement in the state\cite{ckw}. The concept of residual entanglement can be generalized to $N$-party states thus helping in the quantification of 
$N$-party entanglement not accounted for by the bipartite entanglement in its subsystems}.  The nature of monogamy inequality satisfied  by $N$-party states allows one to quantify the $N$-party residual entanglement in addition to shedding light on the extent of limited shareability of entanglement in the state. In view of the fact~\cite{ckw,neg} that different measures of entanglement give rise to different quantifications of the residual entanglement in $3$-qubit pure states, it is natural to expect that similar situation will be realized for $N$-party states also. While it has been shown that generalized (non-symmetric) W states have vanishing concurrence-tangle\cite{ckw} indicating only two-way entanglement in them, they are shown to have non-zero negativity tangle\cite{neg}. It was thereby concluded that the concurrence-tangle underestimates the residual entanglement in $3$-qubit pure states\cite{neg}. 

In this work, we examine the nature of monogamy inequality satisfied  by $N$-qubit pure symmetric states belonging to the W-class. Here the set of all $N$-qubit symmetric states (invariant under the interchange of qubits) with only two distinct qubits (spinors) characterizing them is defined as the W-class of states, owing to the fact that W states form an integral part of it. We show that the monogamy inequality with square of concurrence\cite{con} as the measure of entanglement holds good with equality for all states of this family (quite similar to the behaviour of W states). With squared negativity-of-partial transpose\cite{ppt} as measure of entanglement, we examine the monogamous nature of the W-class of states and show that negativity-tangle has non-zero value for all states in this family. We wish to mention here that the Majorana representation of symmetric $N$-qubit states\cite{maj,bastin,usrmaj} has enabled us to obtain a simplified form for the states with two-distinct spinors thus helping us to obtain the concurrence-tangle and negativity-tangle for the whole family of states.
      
The article is divided into four parts. Section 1  contains introductory remarks. In Section 2, we make use of the Majorana representation of $N$ qubit pure symmetric states to obtain a simplified form of the states belonging to the W class. We analyze the monogamous nature of the W-class of states in Section~3 and evaluate their concurrence-tangle and negativity-tangle.  Section 4  provides a concise summary of the results.

\section{Majorana representation of pure symmetric $N$-qubit states} 
In order to examine the nature of monogamy inequality satisfied by $N$-qubit pure symmetric states of the W-class, we make use of the very elegant Majorana representation~\cite{maj} of pure symmetric states. While several advantages of using the Majorana representation has been reported in the literature\cite{bastin,usrmaj}, we illustrate here its use in identifying the monogamous nature of symmetric states.

In the Majorana representation~\cite{maj}, a pure symmetric state of  $N$ qubits is represented as a {\em symmetrized} combination of $N$ constituent spinors $\vert\epsilon_l\rangle$ as  
\begin{equation}
\label{Maj}
\vert \Psi_{\rm sym}\rangle={\cal N}\, \sum_{P}\, \hat{P}\, \{\vert \epsilon_1, \epsilon_2, 
\ldots  \epsilon_N \rangle\}, 
\end{equation} 
where 
\begin{equation}
\label{spinor}
\vert\epsilon_l\rangle= 
\cos(\beta_l/2)\, e^{-i\alpha_l/2}\, \vert 0\rangle +
\sin(\beta_l/2)\, e^{i\alpha_l/2} \, \vert 1\rangle,\ \ l=0,1,2,\ldots, N.
\end{equation}
Here $\hat{P}$ corresponds to the set of all $N!$ permutations of the spinors (qubits) and ${\cal N}$ corresponds to an overall normalization factor. 

An $N$ qubit pure symmetric state containing $r(<N)$ distinct spinors $\vert \epsilon_i \rangle$ ($i=1,\,2,\,\ldots\, r$), each repeating $n_i$ times,  belongs to the class ${\cal D}_{n_1,\,n_2,\,\ldots\,n_r}$ and each degeneracy configuration $\{n_1,\,n_2,\,\ldots\,n_r \}$ (with the numbers $n_i$ being arranged in the descending order) corresponds to a {\em distinct} SLOCC class\cite{bastin,usrmaj}. The number of SLOCC inequivalent classes possible for states with $r$ distinct spinors is given by the partition function $p(N,\,r)$ that gives the distinct possible ways in which the number  $N$ can be partitioned into $r$ numbers $n_i$  ($i=1,\,2,\,\ldots\, r$) such that $\sum_{i=1}^r \, n_i=N$~\cite{bastin,usrmaj}. For instance, a $3$-qubit state with only one distinct spinor belongs to the class ${\cal D}_3$,  with two distinct spinors belongs to the class $\{{\cal D}_{2,1}\}$ and $\{{\cal D}_{1,1,1}\}$ is the class of $3$-qubit states with three distinct spinors. The classes ${\cal D}_{3}$, ${\cal D}_{2,1}$ and ${\cal D}_{1,1,1}$ are SLOCC inequivalent and a state belonging to one of these classes cannot be converted into the other (different from itself) by any local operations and classical communications\cite{bastin,usrmaj}. While the class ${\cal D}_3$ contains only separable states, $\{{\cal D}_{2,1}\}$ is the W-class of states and $\{{\cal D}_{1,1,1}\}$ corresponds to the GHZ-class of states thus supporting the fact that three qubit pure states can be entangled in two inequivalent ways\cite{dur}. 

A pure symmetric state with $2$ distinct spinors belonging to the SLOCC family 
$\{ {\cal D}_{N-k,k}, k=1,2,\ldots, [N/2]\}$ is given by 
\begin{eqnarray}
\label{dnk}
\vert \Psi_{N-k, k}\rangle &=& {\cal N}\, \sum_{P}\, \hat{P}\,\{ \vert \underbrace{\epsilon_1, \epsilon_1,
\ldots , \epsilon_1}_{N-k};\ \underbrace{\epsilon_2, \epsilon_2,\ldots , \epsilon_2}_{k}\rangle\}\nonumber \\
&=& {\cal N}\, R_1^{\otimes N}\, \sum_{P}\, \hat{P}\,\{ \vert \underbrace{0, 0,
\ldots , 0}_{N-k};\ \underbrace{\epsilon'_2, \epsilon'_2,\ldots , \epsilon'_2}_{k}\rangle\},
\end{eqnarray}
where $\epsilon_1=R_1\vert 0\rangle$ and $\epsilon_2=R_2\vert 0\rangle$, and 
\begin{equation}
\label{ep'}
\vert \epsilon'_2\rangle=R_1^{-1}R_2\vert 0\rangle=d_0\, \vert 0\rangle+d_1\, \vert 1\rangle,\ \ \vert d_0\vert^2+\vert d_1\vert^2=1,\ \ d_1\neq0.
\end{equation}
Thus the symmetric state with two distinct spinors $\vert \Psi_{N-k, k}\rangle$ is shown to be equivalent, up to local unitary transformations, to   
\begin{equation}
\label{nk'}
\vert \Psi_{N-k, k}\rangle \equiv \sum_{r=0}^k\, \sqrt{^N C_{r}}\, \alpha_{r}\, \left\vert\frac{N}{2},\frac{N}{2}-r \right\rangle;\ \ \  
\alpha_{r}={\cal N}\,\, \frac{(N-r)!}{(N-k)! (k-r)!}\, d_0^{k-r}\, d_1^r.
\end{equation}
It can be seen that $\alpha_r=\delta_{k,r}$ when $d_1=1,\ d_0=0$ and the state  $\vert \Psi_{N-k, k}\rangle$ reduces to the Dicke state $\left\vert\frac{N}{2},\frac{N}{2}-k \right\rangle$. It is thus not difficult to see that the states in the family $D_{N-1,1}$ (with $k=1$) are SLOCC equivalent to the $N$-qubit W state $\left\vert\frac{N}{2},\frac{N}{2}-1 \right\rangle$ 

An arbitrary $N$-qubit pure symmetric state belonging to the W-class is  given by 
\begin{eqnarray}
\label{3qu} 
\vert \Psi_{N-1,1} \rangle&=&\sum_{r=0}^1\, \sqrt{^N C_{r}}\, \alpha_{r}\, \left\vert\frac{N}{2},\frac{N}{2}-r \right\rangle =\alpha_0 \left\vert\frac{N}{2},\frac{N}{2} \right\rangle+\sqrt{N} \alpha_1 \left\vert\frac{N}{2},\frac{1}{2} \right\rangle.
\end{eqnarray}
which may be expressed in terms of standard qubit basis as, 
\be
\label{3q2}
\vert \Psi_{N-1,1} \rangle \equiv a \vert 000\cdots 0 \rangle+ b \left(\frac{\vert 100\cdots 0 \rangle+\vert 010\cdots 0 \rangle+\cdots +\vert 00\cdots 01 \rangle}{\sqrt{N}}\right)    
\ee   
with $a=\alpha_0$, $b=\sqrt{N}\,\alpha_1$ are complex numbers obeying $\vert a \vert^2+\vert b \vert^2=1$. On taking $a=\cos \frac{\theta}{2}$, $b=\sin \frac{\theta}{2}\, e^{i\phi}$, ($0<\theta<\pi$, $0<\phi< 2\pi$),  without any loss of generality and subjecting the $N$-qubit state (\ref{3q2})  to another local unitary transformation $\vert 0\rangle'=\vert 0\rangle,\ \ \vert 1\rangle'=e^{-i\phi}\vert 1\rangle$ on all the $N$ qubits we obtain a further simplified form  
\be
\label{sim} 
\vert \Psi_{N-1,1} \rangle \equiv \cos \frac{\theta}{2}\vert 000\cdots 0\rangle  + \sin \frac{\theta}{2} \left(\frac{\vert 100\cdots 0 \rangle+\vert 010\cdots 0 \rangle+\cdots +\vert 00\cdots 01 \rangle}{\sqrt{N}}\right) 
\ee
with a single parameter $\theta$, $0<\theta\leq\pi$ describing the state.    
 
\section{Monogamous nature of pure Symmetric states of the W-class: Concurrence- and negativity- tangle} 
Having obtained the simplified form of the $N$-qubit pure symmetric states with two distinct spinors, we will use it to evaluate the concurrence- and negativity tangle of this family and thereby make a statement about their monogamous nature with respect to different entanglement measures. We carry out this task in the following. 

\subsection{Concurrence-tangle:}
We start by recalling the monogamy inequality in terms of squared-concurrence in three-qubit systems introduced by Coffman, Kundu and Wootters (CKW)~\cite{ckw}. They\cite{ckw} have shown that for any $3$-qubit pure state $\Psi_{ABC}$,
\be
C^2_{AB} + C^2_{AC} \leq C^2_{A:BC}
\ee
where $C_{AB} ( C_{AC})$ is the concurrence between $A$, $B$ ($C$), while $C_{A:BC}=2\, \sqrt{\mbox{det}\,\rho_A}$ is the concurrence between system $A$ and $BC$. 
The quantity $C^2_{A:BC}-(C^2_{AB} + C^2_{AC})$ is known as {\emph {three-tangle}} or {\emph {concurrence-tangle}}  and is a measure of three-party entanglement\cite{ckw}. It was also conjectured\cite{ckw} that a monogamy relation of the form  
\be
C_{A_{1}A_{2}}^{2} + C_{A_{1}A_{3}}^{2}+C_{A_{1}A_{4}}^{2} + \cdots+ C_{A_{1}A_{n}}^{2} \leq C_{A_{1}:A_{2}A_{3}A_{4} \ldots A_{N}}^{2}
\ee
holds good for all $N$-qubit pure states. We can term the quantity 
\be
C_{A_{1}:A_{2}A_{3}A_{4} \ldots A_{N}}^{2}-\left( C_{A_{1}A_{2}}^{2} + C_{A_{1}A_{3}}^{2}+C_{A_{1}A_{4}}^{2} + \cdots+ C_{A_{1}A_{N}}^{2} \right)
\ee
as $N$-{\emph{concurrence-tangle}}. In fact, it was shown in Ref. ~\cite{ckw} that generalized (non-symmetric) $3$-qubit W states given by 
$a\vert 100 \rangle+b\vert 010\rangle+c\vert 001 \rangle$ have vanishing concurrence-tangle and indicated that their $N$-qubit counterparts will also exhibit the same feature. We wish to illustrate here that all pure symmetric $N$-qubit states with two-distinct spinors, the W-class of states, have  vanishing N-concurrence-tangle. Towards this end we first wish to evaluate the form of the two-qubit and single-qubit reduced density matrices of the $N$-qubit state $\vert \Psi_{N-1,1} \rangle$.  Knowing the structure of single qubit density matrices is essential to obtain $C_{A_{1}:A_{2}A_{3}A_{4} \ldots A_{N}}=2 \, \sqrt{\mbox{det}\,\rho_A}$~\footnote{Though the concurrence is defined only for two-qubit systems, the $N$ qubit state being pure the $N-1$-qubits essentially belong to a two-dimensional space and hence one can define the concurrence between a single qubit and the remaining $N-1$ qubits\cite{ckw}. Also, the effective concurrence is the concurrence between two-qubits in a pure state leading to $C_{A_{1}:A_{2}A_{3}A_{4} \ldots A_{N}}=2 \, \sqrt{\mbox{det}\,\rho_A}$.}, the structure of two-qubit (mixed) density matrices is needed for the evaluation of $C_{A_{1}A_{2}}$. We need to note here that  $\vert \Psi_{N-1,1} \rangle$ being a symmetric state, all its two-qubit and single-qubit subsystems are identical, irrespective of which two qubits or single qubit we choose to consider. That is, 
\begin{eqnarray}
\rho_{A_{1}A_{2}}&=& \rho_{A_{1}A_{3}}=\rho_{A_{2}A_{3}}=\cdots=\rho_{A_{N-1}A_{N}} \nonumber \\
\rho_{A_{1}}&=& \rho_{A_{2}}=\rho_{A_{3}}=\cdots=\rho_{A_{N}}
\end{eqnarray}

The form of the single-qubit, two-qubit marginals of the state $\vert \Psi_{N-1,1} \rangle$ for $N=3,\,4,\,5,\,6$ allows us to generalize and obtain these marginals for any $N$. 
In Table~1, we have tabulated the structure of reduced density matrices $\rho_{A_{1}A_{2}}$, $\rho_{A_{1}}$ of $\vert \Psi_{N-1,1} \rangle$. 

\begin{table}[ht] 
\begin{center} 
\caption{The single qubit and two-qubit marginals of $\vert \Psi_{N-1,1} \rangle$  for $N=3\ to \ 6$}
\textbf{
\begin{tabular}{|c|c|c|}
\hline 
& & \\ 
$N$ & $\rho_{A_{1}A_{2}}$ & $\rho_{A_1}$  \\ & & \\
\hline\hline & & \\ 
3 &  $\frac{1}{6}\ba{cccc} 2(2+\cos \theta) & \sqrt{3}\sin \theta & \sqrt{3}\sin \theta & 0 \\ \sqrt{3}\sin \theta & 1-\cos \theta & 1-\cos \theta & 0 \\ \sqrt{3}\sin \theta & 1-\cos \theta & 1-\cos \theta & 0 \\0 & 0 & 0 & 0 \ea$ &  $\frac{1}{6}\ba{cc} 5+\cos \theta & \sqrt{3} \sin \theta \\ \sqrt{3} \sin \theta & 1-\cos \theta \ea$  \\ & &  \\ \hline  
& &  \\ 
4 &  $\frac{1}{8}\ba{cccc} 2(3+\cos \theta) & 2\sin \theta & 2\sin \theta & 0 \\ 2 \sin \theta & 1-\cos \theta & 1-\cos \theta & 0 \\ 2 \sin \theta & 1-\cos \theta & 1-\cos \theta & 0 \\0 & 0 & 0 & 0 \ea$ & $\frac{1}{8}\ba{cc} 7+\cos \theta & 2 \sin \theta \\ 2 \sin \theta & 1-\cos \theta \ea $ \\ & &  \\ \hline  
& & \\ 
5 & $ \frac{1}{10}\ba{cccc} 2(4+\cos \theta) & \sqrt{5}\sin \theta & \sqrt{5}\sin \theta & 0 \\ \sqrt{5}\sin \theta & 1-\cos \theta & 1-\cos \theta & 0 \\ \sqrt{5}\sin \theta & 1-\cos \theta & 1-\cos \theta & 0 \\0 & 0 & 0 & 0 \ea $ & $\frac{1}{10}\ba{cc} 9+\cos \theta & \sqrt{5} \sin \theta \\ \sqrt{5} \sin \theta & 1-\cos \theta \ea$ \\ & &  \\ \hline 
& & \\  
6 & $\frac{1}{12}\ba{cccc} 2(5+\cos \theta) & \sqrt{6}\sin \theta & \sqrt{6}\sin \theta & 0 \\ \sqrt{6}\sin \theta & 1-\cos \theta & 1-\cos \theta & 0 \\ \sqrt{6}\sin \theta & 1-\cos \theta & 1-\cos \theta & 0 \\0 & 0 & 0 & 0 \ea $ & $ \frac{1}{12}\ba{cc} 11+\cos \theta & \sqrt{6} \sin \theta \\ \sqrt{6} \sin \theta & 1-\cos \theta \ea$ \\ & &  \\ \hline 
\end{tabular}}
\end{center}
\end{table} 
Using the form of two-qubit and single-qubit density matrices given in Table~1, we can readily obtain the structure of the two-qubit and single-qubit density matrices of the $N$-qubit state $\vert \Psi_{N-1,\,1}\rangle$ for any $N$. We have
\be 
\label{2qgen}
\rho_{A_{1}A_{2}}=\frac{1}{2N}\ba{cccc} 2(N-1+\cos \theta) & \sqrt{N}\sin \theta & \sqrt{N}\sin \theta & 0 \\ \sqrt{N}\sin \theta & 1-\cos \theta & 1-\cos \theta & 0 \\ \sqrt{N}\sin \theta & 1-\cos \theta & 1-\cos \theta & 0 \\0 & 0 & 0 & 0 \ea 
\ee
and
\begin{eqnarray}
\label{1qgen}
\rho_{A_{1}}&=&\frac{1}{2N}\ba{cc} 2N-1+\cos \theta & \sqrt{N} \sin \theta \\ \sqrt{N} \sin \theta & 1-\cos \theta \ea
\end{eqnarray}
The concurrence\cite{con} of the two-qubit state $\rho_{A_{1}A_{2}}$ is given by $C_{A_{1}A_{2}}=\mbox{max}\left(0,\,\sqrt{\lambda_1}-\sqrt{\lambda_2}-\sqrt{\lambda_3}-\sqrt{\lambda_4} \right)$ where $\lambda_i$, $i=1,\,2,\,3,\,4$ are the eigenvalues of the  matrix $\rho_{A_{1}A_{2}}\rho'_{A_{1}A_{2}}$, $\rho'_{A_{1}A_{2}}=({\sigma}_{y}\otimes{\sigma}_{y}) \rho^{*}_{A_{1}A_{2}} ({\sigma}_{y}\otimes{\sigma}_{y})$ being the spin-flipped density matrix. It can be seen that there is only one non-zero eigenvalue   $\lambda=\frac{1}{N^{2}}(1-\cos \theta)^2$ for $\rho_{A_{1}A_{2}}\rho'_{A_{1}A_{2}}$ and we therefore have \footnote{Notice here that the concurrence between any two qubits becomes maximum and is equal to $2/N$ when $\theta=\pi$ for the W-states. In fact this is the maximum bipartite entanglement in an $N$-qubit state, achievable for W-states, as is shown in Ref. ~\cite{kw1}.} 
\be
C_{A_{1}A_{2}}=C_{A_{1}A_{3}}=\cdots=C_{A_{1}A_{N}}=\sqrt{\lambda}=\frac{1}{N}(1-\cos \theta).
\ee 
Similarly, we obtain $\det \rho_{A}=\frac{N-1}{4N^{2}}(1-\cos \theta)^2$ and hence  
\be
\label{1qgenf}
C^{2}_{A_{1}:A_{2}A_{3}\ldots A_{N}}=4 \det (\rho_{A})=\frac{N-1}{N^{2}}(1-\cos \theta)^2. 
\ee 
As there are $N-1$ identical two-qubit subsystem density matrices $\rho_{A_{1}A_{i}}$, $i=2,\,3,\,\ldots N$, with the first qubit being common to all of them, we have  
\be
C_{A_{1}A_{2}}^{2} + C_{A_{1}A_{3}}^{2}+C_{A_{1}A_{4}}^{2} + \cdots+ C_{A_{1}A_{N}}^{2}=(N-1)C_{A_{1}A_{2}}^{2}=\frac{N-1}{N^2}(1-\cos \theta)^2
\ee
Now, we can readily see that  (See Eq.(\ref{1qgenf}))
\be
C_{A_{1}A_{2}}^{2} + C_{A_{1}A_{3}}^{2}+C_{A_{1}A_{4}}^{2} + \cdots+ C_{A_{1}A_{N}}^{2}=\frac{N-1}{N^2}(1-\cos \theta)^2=C_{A_{1}:A_{2}A_{3}A_{4} \ldots A_{N}}^{2}
\ee
establishing the relation
\be
\label{mon}
C_{A_{1}A_{2}}^{2} + C_{A_{1}A_{3}}^{2}+C_{A_{1}A_{4}}^{2} + \cdots+ C_{A_{1}A_{N}}^{2} = C_{A_{1}:A_{2}A_{3}\ldots A_{N}}^{2}
\ee
for the $N$-qubit pure states of the W-class. 
Thus in addition to verifying the monogamy inequality, we have shown that \emph{equality} holds good for all $N$-qubit states belonging to the W-class. In other words, we have shown that the N-concurrence tangle 
$C_{A_{1}:A_{2}A_{3}\ldots A_{N}}^{2}-(N-1)C_{A_{1}A_{2}}^{2}$ vanishes for the family of states $\vert \Psi_{N-1,1} \rangle$.  

\subsection{Negativity tangle:} 
We begin here by recalling that a monogamy inequality for $3$-qubit pure states in terms of negativity-of-partial transpose has been proposed in Ref. ~\cite{neg}. While the vanishing concurrence-tangle for W-states indicated only two-way entanglement, the analogous quantity defined by~\cite{neg} 
$\Pi_A=N^2_{A:BC}-N^2_{AB}-N^2_{AC}$ showed a non-zero three-way entanglement in W states. 
While the concurrence-tangle is independent of the focus qubit, the negativity tangle depends on which qubit is considered as the focus qubit. Thus the negativity tangle for $3$-qubit pure states is defined as 
$\Pi=\frac{1}{3}(\Pi_A+\Pi_B+\Pi_C)$ with $\Pi_A$, $\Pi_B$, $\Pi_C$ being the negativity tangles with the focus qubits being $A$, $B$, $C$ respectively. 
  
Quite similarly to the case of $3$-qubit pure states, the negativity tangle for $N$-qubit pure states is defined as\cite{neg}
\bea
\Pi&=&\frac{1}{N}(\Pi_1+\Pi_2+\ldots+\Pi_N)\ \ \mbox{where} \\
\Pi_{1}&=&N^{2}_{A_{1}:A_{2}A_{3}\ldots A_N}-\left(N^{2}_{A_{1}A_{2}}+N^{2}_{A_{1}A_{3}}+\ldots+N^{2}_{A_{1}A_{N}}\right) \nonumber \\
\Pi_{2}&=&N^{2}_{A_{2}:A_{1}A_{3}\ldots A_N}-\left(N^{2}_{A_{2}A_{1}}+N^{2}_{A_{2}A_{3}}+\ldots+N^{2}_{A_{2}A_{N}}\right) \ \nonumber \\
\vdots &=& \vdots \nonumber \\
\Pi_{N}&=&N^{2}_{A_{N}:A_{1}A_{2}\ldots A_{N-1}}-\left(N^{2}_{A_{N}A_{1}}+N^{2}_{A_{N}A_{2}}+\ldots+N^{2}_{A_{N}A_{N-1}}\right) \nonumber 
\eea       
While Ref. ~\cite{ckw} indicated vanishing concurrence tangle for $N$-qubit generalized (non-symmetric) W states, Ref. ~\cite{neg} illustrated that they have a residual $N$-party entanglement quantified by $\Pi$, the negativity tangle. Here, we show that the whole family of pure $N$-qubit symmetric states belonging to two-distinct spinors (the W-class of states) have non-zero residual entanglement when quantified through negativity tangle. We illustrate this aspect in the following.

Having obtained the two-qubit reduced density matrices of the symmetric state $\vert \Psi_{N-1,1}\rangle$ (See Eq. (\ref{2qgen})), we can readily evaluate their negativity of partial transpose\cite{ppt}.  The partially transposed density matrix of the two-qubit reduced density matrix $\rho_{A_{1}A_{2}}$ obtained in Eq. (\ref{2qgen}) is evaluated to be
\be\rho^T_{A_{1}A_{2}}=\frac{1}{2N}\ba{cccc} 2(N-1+\cos \theta) & \sqrt{N}\sin \theta & \sqrt{N}\sin \theta & 1-\cos \theta \\ \sqrt{N}\sin \theta & 1-\cos \theta & 0 & 0 \\ \sqrt{N}\sin \theta & 0 & 1-\cos \theta & 0 \\1-\cos \theta & 0 & 0 & 0 \ea
\ee
The negativity of partial transpose is given by $(\vert\vert \rho^T_{A_{1}A_{2}} \vert \vert-1)/2$ where $\vert\vert \rho^T_{A_{1}A_{2}} \vert \vert$ is the tracenorm of the partially transposed density matrix $\rho^T_{A_{1}A_{2}}$ and it is the sum of the square-root of eigenvalues of the positive-definite matrix $\left(\rho^T_{A_{1}A_{2}}\right)^\dag \rho^T_{A_{1}A_{2}}$. As the negativity for a two-qubit system varies from $0$ to $0.5$, we choose to take $N_{A_{1}A_{2}}$ to be   
\be
N_{A_{1}A_{2}}=\vert\vert \rho^T_{A_{1}A_{2}} \vert \vert-1 
\ee
so that it varies from $0$ to $1$, quite similar to the variation of concurrence. In fact, this is the convention adopted for negativity in Ref. ~\cite{neg} while obtaining the negativity tangle for three-qubit pure states.

As the negativity of a permutation invariant state is identical for any pair of qubits, we denote $N_{A_{1}A_{2}}=N_{A_{1}A_{k}}$, $k=2,\,3,\,\ldots,\,N$. Fig.1 shows the plot of negativity $N_{A_{1}A_{k}}$ with respect to $\theta$ for the W-class of states $\vert \Psi_{N-1,1} \rangle$.
\begin{figure}[ht]
\centerline{\includegraphics* [width=2.4in,keepaspectratio]{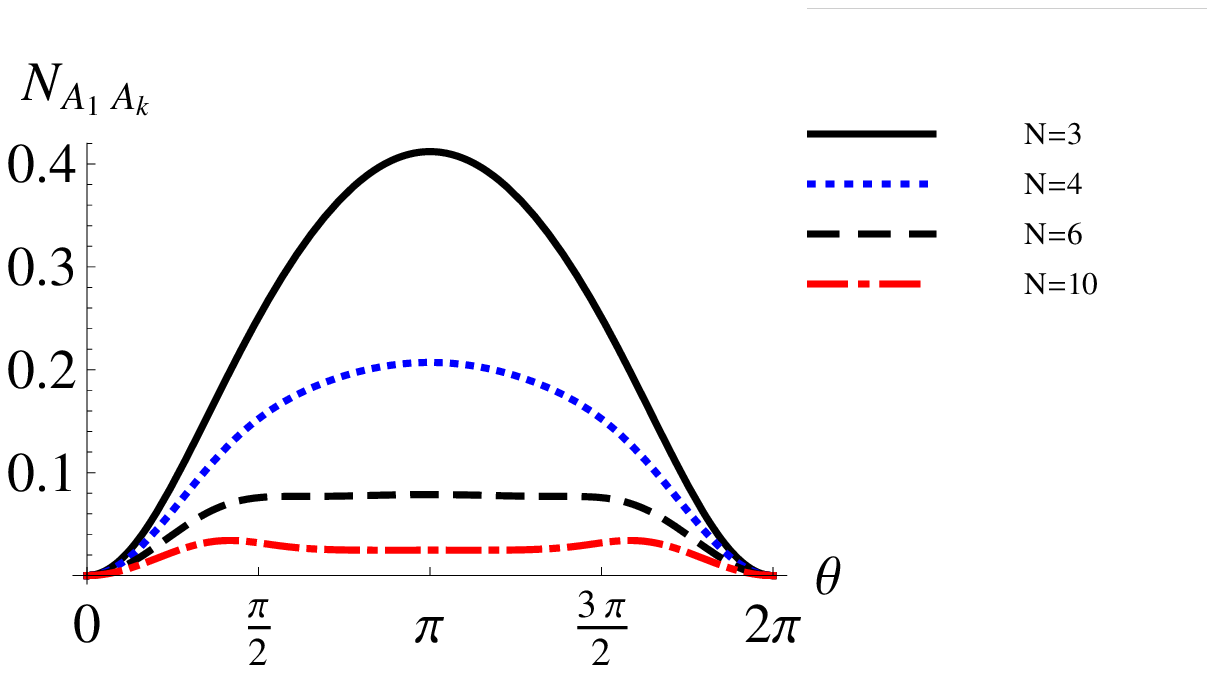}} 
\caption{The plot of $N_{A_{1}A_{k}}$, versus $\theta$ in the interval $0$ to $2\pi$ for arbitrary $N$ qubit state belonging to the W-class.}  
\end{figure}

It can be seen that with the increase in the number of qubits, the pairwise entanglement quantified by negativity of partial transpose $N_{A_{1}A_{k}}$ decreases quite considerably. 

In Ref. ~\cite{neg} it was shown that the negativity between the focus qubit and the remaining two qubits of a pure $3$-qubit state matches with their concurrence.  The same argument can be extended to $N$-qubit pure states yielding\footnote{The authenticity of the relation $N_{A_{1}:A_{2}A_{3}\ldots A_{N}}=C_{A_{1}:A_{2}A_{3}\ldots A_{N}}$ is explicitly verified for $N=3,\,4,\,5,\,6$ and generalized thereby to any $N$.} 
\be
N_{A_{1}:A_{2}A_{3}\ldots A_{N}}=C_{A_{1}:A_{2}A_{3}\ldots A_{N}}=2\sqrt{\mbox{det}\,\rho_{A_1}}.
\ee
The variation of $N_{A_{1}:A_{2}A_{3}\ldots A_{N}}$ with $\theta$, for different values of $N$, is as shown in Fig.~2. 
\begin{figure}[ht]
\centerline{\includegraphics* [width=2.4in,keepaspectratio]{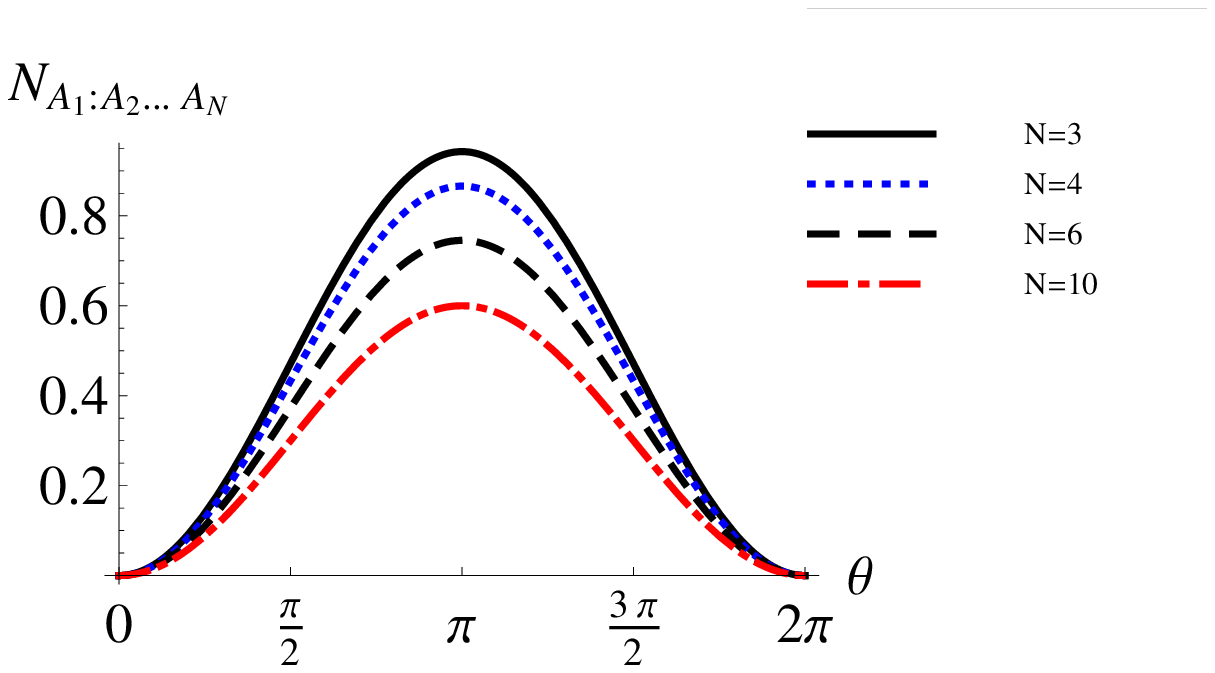}}
\caption{The plot of $N_{A_{1}:A_{2}A_{3}\ldots A_{N}}$, versus $\theta$ in the interval $0$ to $2\pi$ for arbitrary $N$ qubit state belonging to the W-class.}  
\end{figure}
With $\mbox{det}\,\rho_{A_1}$ being $\frac{N-1}{4N^{2}}(1-\cos \theta)^2$, we have the negativity tangle $\Pi_1$ as  
\bea
\Pi_{1}&=&N^{2}_{A_{1}:A_{2}A_{3}\ldots A_N}-\left(N^{2}_{A_{1}A_{2}}+N^{2}_{A_{1}A_{3}}+\ldots+N^{2}_{A_{1}A_{N}}\right)  \nonumber \\ 
&=&  \frac{N-1}{N^{2}}(1-\cos \theta)^2-(N-1)N^{2}_{A_{1}A_{2}}
\eea
But as we are considering symmetric states that are invariant under permutation of qubits, $\Pi_1=\Pi_2=\ldots=\Pi_N$ and hence, 
\bea
\Pi_w&=&\frac{\Pi_{1}+\Pi_{2}+\ldots +\Pi_{N}}{N}=\Pi_1 \nonumber \\ 
&=&4\,\mbox{det}\,\rho_{A_1}-(N-1)N^{2}_{A_{1}A_{k}}=(N-1)\left(\frac{(1-\cos \theta)^2}{N^{2}}-N^{2}_{A_{1}A_{k}}\right)  
\eea
is the negativity tangle of the state $\vert \Psi_{N-1,1}\rangle$ belonging to the W-class. We plot a graph of negativity tangle $\Pi_w$ as a function of $\theta$ in Fig 3.\\
\begin{figure}[ht]
\centerline{\includegraphics* [width=2.4in,keepaspectratio]{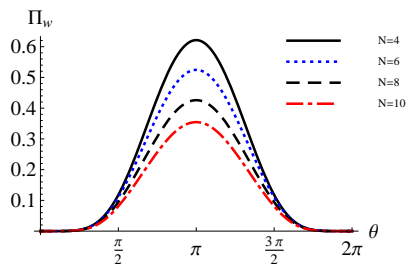}} 
\caption{The plot of negativity tangle $\Pi_W$ versus $\theta$ for the $N$-qubit symmetric state belonging to the W-class.} 
\end{figure}  

In particular, for $N$-qubit W-states, the negativity-tangle is given by 
\be
\Pi_w=\frac{N-1}{N^2}\left(4-\left[\sqrt{(N-2)^2+4}-(N-2)\right]^2\right) 
\ee
Fig. 4 shows the variation of negativity tangle with the number of qubits $N$ for $N$-qubit W states (corresponding to $\theta=\pi$ in $\vert \Psi_{N-1,1} \rangle$.) 
\begin{figure}[ht]
\centerline{\includegraphics*[width=2.4in,keepaspectratio]{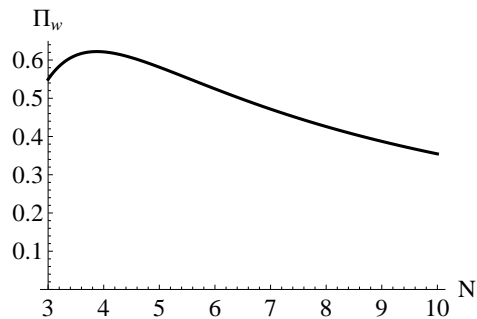}} 
\caption{The negativity-tangle of $N$-qubit W states as a function of number of qubits $N$. It can be seen that $\Pi_{w}$ is maximum when $N=4$ and decreases with the number of qubits.} 
\end{figure} 

We have thus accomplished the task of evaluating the negativity-tangle for $N$-qubit pure states 
belonging to the W-class and illustrated the fact that the concurrence-tangle underestimates 
the residual entanglement in $N$-qubit states also. In addition, we have shown that the 
negativity-tangle which quantifies the residual entanglement in the $N$-qubit states 
decreases with increase in $N$ for $N\geq 4$. In fact, as can be seen from Figs. 3 and 4, the 
three-qubit states belonging to the W-class have lesser residual entanglement than their four-qubit 
counterparts and for $N\geq 4$, the negativity-tangle goes on decreasing monotonically. Also, one can 
observe that though the  
bipartite entanglement $N_{A_{1}A_{k}}$ (See Fig. 1) decreases quite drastically with increase in the number of qubits, the decrease in the residual entanglement $\Pi_w$ with $N$ is relatively smaller (See Fig. 3). This is due to the slower decrease of the $1:N-1$ entanglement, quantified through $N_{A_{1}:A_{2}A_{3}\ldots A_N}$, with the increase of qubits (as compared to the fast decrease of $N_{A_{1}A_{i}}$ with $N$)(See Figs 1 and 2).      
\section{Conclusion} 
In this article, we have analyzed the monogamous nature of $N$-qubit pure states belonging to the W-class using squared concurrence and squared negativity as measures of bipartite entanglement. 
Using the simplified form of the states belonging to the W-class, obtained using the Majorana representation of $N$-qubit symmetric pure states, we have evaluated the $N$-concurrence-tangle and negativity-tangle of this family of states. Quite similar to the $N$-qubit W-states, we have shown that all states in the W-class of states have vanishing concurrence-tangle. By showing that W-class of states have non-zero negativity-tangle, we have proved the fact that concurrence-tangle underestimates the residual entanglement even in $N$-qubit states with $N\geq 3$. It would be of interest to examine the nature of monogamy inequality in $N$-qubit symmetric states belonging to different SLOCC inequivalent families and compare their residual entanglement.    

\section*{Acknowledgments}
P.J. Geetha  acknowledges the support of Department of Science and Technology (DST), Govt. of India through the award of INSPIRE fellowship.

\end{document}